\documentclass[runningheads]{llncs}
\usepackage{amssymb}
\usepackage[T1]{fontenc}
\usepackage{graphicx}
\usepackage{hyperref}
\usepackage{multirow}
\usepackage{multicol}
\usepackage{subcaption}
\usepackage{appendix}
\usepackage[misc]{ifsym}
\begin{document}
\title{Leave No One Behind: Enhancing Diversity While Maintaining Accuracy in Social Recommendation}
\titlerunning{Enhancing Diversity While Maintaining Accuracy in Social Recommendation}
%
\author{Lei Li \and
Xiao Zhou \textsuperscript{(\Letter)} \protect\footnotetext{\textsuperscript{\Letter} Corresponding Author}}

\authorrunning{L. Li and X. Zhou}
%
\institute{Gaoling School of Artificial Intelligence, Renmin University of China, Beijing, China
\email{\{leil, xiaozhou\}@ruc.edu.cn}}
\maketitle              
\begin{abstract}
Social recommendation, which incorporates social connections into recommender systems, has proven effective in improving recommendation accuracy. However, beyond accuracy, diversity is also crucial for enhancing user engagement. Despite its importance, the impact of social recommendation models on diversity remains largely unexplored. In this study, we systematically examine the dual performance of existing social recommendation algorithms in terms of both accuracy and diversity. Our empirical analysis reveals a concerning trend: while social recommendation models enhance accuracy, they often reduce diversity. To address this issue, we propose \textbf{Diversified Social Recommendation (DivSR)}, a novel approach that employs \textbf{relational knowledge distillation} to transfer high-diversity structured knowledge from non-social recommendation models to social recommendation models. DivSR is a lightweight, model-agnostic framework that seamlessly integrates with existing social recommendation architectures. Experiments on three benchmark datasets demonstrate that DivSR significantly enhances diversity while maintaining competitive accuracy, achieving a superior accuracy-diversity trade-off. Our code and data are publicly available at: \url{https://github.com/ll0ruc/DivSR}.

\end{abstract}
\section{Introduction}
In the era of information overload, recommender systems play a crucial role in helping users navigate vast amounts of content~\cite{davidson2010youtube}. They have been successfully deployed across various domains, including e-commerce~\cite{schafer2001commerce}, online news~\cite{yang2016effects}, and multimedia content~\cite{park2012personalized}. With the advancement of recommendation algorithms, accuracy has become the primary, and often sole, optimization objective~\cite{amatriain2009rate,kang2018self}.  As a major branch of recommender systems, social recommendation~\cite{ma2008sorec,tao2022revisiting} leverages social resources, such as interpersonal relationships and influence, to enhance recommendation performance. This approach typically incorporates social connections either as a regularization constraint~\cite{jamali2010matrix} or by extracting feature embeddings from neighboring nodes using a graph attention network framework~\cite{fan2019graph}.  

However, an accurate recommendation is not necessarily a satisfactory one~\cite{avazpour2014dimensions}. Users on e-commerce platforms seek more than just highly relevant products; they also devote significant time to exploring news-feed products for a wider range of options. Thus, an ideal recommendation system should fulfill both accuracy and diversity requirements~\cite{parapar2021towards}. Unfortunately, existing social recommendation systems primarily focus on enhancing accuracy, often overlooking diversity. This oversight is problematic considering that social recommendation systems are fundamentally influenced by social influence theory~\cite{marsden1993network}, which posits that users influenced by their social connections tend to adopt similar preferences. This can lead to homogeneity in the recommendations over time due to the over-reliance on preferences within a user's immediate social circle~\cite{sacharidis2020fairness}, underlining the necessity of fostering diversity in social recommendations. 

In this work, we revisit current social recommendation algorithms, probing their performance in terms of diversity. We conduct extensive experiments to evaluate the performance of several existing social recommendation methods on three benchmark datasets assessing both accuracy and diversity. For each algorithm, we remove socially-relevant modules, enabling it to make item recommendations without utilizing social information. By comparing social recommendation methods with their non-social variants, we can identify performance discrepancies resulting from the integration of social relationships. Our empirical findings reveal that \textit{existing social recommendation models tend to decrease diversity while improving accuracy compared to non-social recommendation methods}.

To achieve a better accuracy-diversity trade-off in social recommendation, we propose the DivSR (\textbf{Div}ersified \textbf{S}ocial \textbf{R}ecommendation) framework, which leverages knowledge distillation to achieve system-level overall diversity in recommendations. Fundamentally, DivSR maintains a model-agnostic design, allowing seamless integration with various social recommendation backbone models. In DivSR, we train a social recommendation model as the student model to combine high accuracy and high diversity. The high diversity is derived from a pre-trained teacher model, which serves as a non-social recommendation counterpart. We design a knowledge transfer module using relational distillation learning technology~\cite{park2019relational}, which distills structured similarity knowledge between users and their social connections from the teacher model to the student model. To strike a balanced trade-off between accuracy and diversity, DivSR optimizes both the recommendation task and the knowledge distillation task simultaneously within a primary and auxiliary learning framework. 

We perform experiments on three widely used public datasets, incorporating five robust social recommenders as backbone models. Comprehensive results demonstrate that DivSR enhances diversity without significantly sacrificing accuracy across various social recommendation backbones. DivSR achieves a superior accuracy-diversity trade-off compared to several diversified models. The key contributions of this paper are summarized as follows:

\begin{itemize}
\item We empirically assess the accuracy-diversity performance of social recommendation systems and find that they typically reduce diversity compared to their non-social counterparts.

\item We propose DivSR, a model-agnostic framework that leverages knowledge distillation techniques to foster diversity in social recommendations. This approach includes a diversity-knowledge transfer module that distills structured similarity information.

\item Through experiments on three datasets, we demonstrate the effectiveness of DivSR, highlighting its ability to significantly enhance diversity without significantly sacrificing accuracy across various social recommendation systems.
\end{itemize}

\section{Related Work}

\subsection{Social Recommendation}
To improve the accuracy of recommendation results, numerous social recommendation methods have been developed, incorporating online social relationships between users as side information~\cite{guo2015trustsvd,li2021spex}. Early models like SoReg~\cite{ma2011recommender} and SocialMF~\cite{jamali2010matrix} integrate social connections as regularization terms or utilize trust relationships to project users into latent representations. In recent years, graph neural networks (GNNs)~\cite{kipf2016semi} have achieved great success in deep learning, owing to their strong capability on modeling graph data. DiffNet~\cite{wu2019neural} and its extension DiffNet++~\cite{wu2020diffnet++} model the information diffusion process in social graphs to enlarge the users’ influence scope. Multi-channel hypergraph convolutional network is employed on MHCN~\cite{yu2021self} to enhance social recommendation by leveraging high-order user relations. Self-supervised learning (SSL) is utilized in SEPT~\cite{yu2021socially} to improve social recommendation by uncovering supervisory signals from two complementary views of raw data. DESIGN~\cite{tao2022revisiting} introduces knowledge distillation between models that rely on different data sources to leverage social information effectively. A universal denoised self-augmented learning framework (DSL)~\cite{wang2023denoised} incorporates social influence to decipher user preferences while mitigating noisy effects. Nevertheless, these methods mostly aim to improve accuracy while neglecting diversity. Our work contributes to achieving a balance between accuracy and diversity in social recommendation.
 
\subsection{Diversified Recommendation}
Diversified recommendation aims to provide users with a more varied set of items, enabling users to discover new and unexplored interests~\cite{chen2020improving,liang2021enhancing}. The accuracy-diversity dilemma, pointing higher accuracy often means losing diversity to some extent and vice versa. A classical re-ranking work to enhance diversity is maximal marginal relevance (MMR)~\cite{carbonell1998use}, which uses the notion of marginal relevance to combine relevance and diversity with a trade-off parameter. Determinantal point process (DPP)~\cite{chen2018fast} re-ranks items to achieve the largest determinant on the item’s similarity matrix. DGCN~\cite{zheng2021dgcn} selects node neighbors based on the inverse category frequency for diverse aggregation and further utilizes category-boosted negative sampling and adversarial learning to diverse items in the embedding space. DGRec~\cite{yang2023dgrec} targets diversifying GNN-based recommender systems with diversified embedding generation. Different from these methods, our work facilitates seamless integration with various social recommendation systems, effectively enhancing accuracy while maintaining high diversity.

\section{PRELIMINARIES}
\label{sec:section3}
\subsection{Problem Statement}
In the task of social recommendation, let $\textit{U}=\{u_1,u_2,\ldots,u_M\}$ (|\textit{U}| = \textit{M}) represent the set of users, and $\textit{V} =\{v_1,v_2,\ldots, v_N\}$ (|\textit{V}| = \textit{N}) represent the set of items. We use indices $a$ and $b$ to refer to users, and $i$ and $j$ to refer to items. Let $G_s=<U, S>$ denote a directed social graph, where $S \in \mathbb{R}^{M\times M}$ is a matrix representing social relations between users. For each user-user pair ($a,b$), $s_{ab}$ = 1 if user $a$ trusts user $b$ and 0 otherwise. Let $G_r=<U \cup V, R>$ denote an undirected bipartite graph, where $R \in \mathbb{R}^{M\times N}$ is a user-item interaction matrix. For each user-item pair ($a,i$), $r_{ai}$ = 1 indicates that user $a$ has interacted with item $i$ and 0 otherwise. The primary objective of a social recommendation system leveraging $G_r$ and $G_s$ is to predict and recommend the top $k$ items a user is likely to be interested in, based on their past interactions and social influences. The diversified social recommendation task aims to recommend items that users prefer while ensuring high system-level overall diversity.

\subsection{Accuracy-Diversity Dilemma in Social Recommendation}
\label{sec:Acc-Div}
In this paper, we revisit existing social recommendation algorithms, probing their performance in terms of accuracy and diversity. Nevertheless, the comparison of model performance across different social recommendation approaches is not the primary focus of this study. Instead, we specifically examine how the incorporation of social relationships influences both accuracy and diversity. We conduct preliminary experiments to evaluate the two-fold performance \textit{w.r.t.} accuracy (Recall@100) and diversity (Coverage@100) between Social Recommender System (\textbf{Social RS}) and Non-social Recommender System (\textbf{Non-social RS}). These experiments are conducted on three widely-used public datasets (Yelp~\cite{wu2019neural}, Ciao~\cite{tang2012mtrust}, and Flickr~\cite{wu2019hierarchical}). 

\begin{table}
\centering
  \caption{Accuracy-Diversity results on Yelp, Ciao, and Flickr datasets.}
  \label{tab:TRADE-OFF}
  \setlength{\tabcolsep}{6pt}
  \begin{tabular}{l|cc|cc|cc}
  \hline
     &\multicolumn{2}{|c|}{Yelp}  &\multicolumn{2}{|c}{Ciao} &\multicolumn{2}{|c}{Flickr}\\
      &Recall  &Coverage &Recall  &Coverage &Recall  &Coverage  \\
     \hline
  $\sim$ (w/o social) &12.513 &31.610 &12.301 &33.478 &2.219 &40.723 \\
  TrustMF &13.773 $\uparrow$ &23.276$\downarrow$ &12.778$\uparrow$ &21.524$\downarrow$ &3.242$\uparrow$ &21.547$\downarrow$\\
    \hline
   $\sim$ (w/o social) &12.513 &31.610 &12.301 &33.478 &2.219 &40.723 \\
    SocialMF  &14.077$\uparrow$ &21.335$\downarrow$ &12.899$\uparrow$ &22.215$\downarrow$ &3.339$\uparrow$ &27.125 $\downarrow$\\
    \hline
    $\sim$ (w/o social) &12.543 &39.974 &12.519 &39.655 &2.135 &43.654 \\
    DiffNet &14.136$\uparrow$ &16.738$\downarrow$ &12.658$\uparrow$ &27.560$\downarrow$ &3.539$\uparrow$ &32.663$\downarrow$ \\
    \hline
   $\sim$ (w/o social)  &14.518 &59.502 &15.803 &19.746 &3.152 &22.905 \\
    MHCN &15.365$\uparrow$ &53.273$\downarrow$ &16.338$\uparrow$ &22.173$\uparrow$ &4.574 $\uparrow$ &40.205$\uparrow$ \\
    \hline
    $\sim$ (w/o social)  &13.553 &51.277 &14.102 &56.267 & 3.427 & 32.761\\
    DESIGN  &14.984$\uparrow$ &42.228$\downarrow$ &15.367$\uparrow$ &24.262$\downarrow$ &4.127$\uparrow$ &35.290$\uparrow$ \\
   \hline
   \end{tabular}
\end{table}

For Social RS, we select several representative social recommendation models, including TrustMF~\cite{yang2016social}, SocialMF~\cite{jamali2010matrix}, DiffNet~\cite{wu2019neural}, DESIGN~\cite{tao2022revisiting}, and MHCN~\cite{yu2021self}. In contrast, to create the non-social recommendation counterpart, we remove the socially relevant modules from each Social RS while keeping all other components intact. For instance, DiffNet utilizes a layer-wise GNN structure to simulate the recursive social diffusion process. The final user embedding $p_a$ consists of two parts: the embedding derived from social diffusion layers and the preferences based on historical interactions:
\begin{equation}
  p_a = h_a^K + \sum_{i\in \mathcal{N}_r(a)}\frac{q_i}{\left| \mathcal{N}_r(a) \right|},
\end{equation}
\begin{equation}
  h_a^K = GNN(h_a^0;G_s),
\end{equation}
where $h_a^0$ and $q_i$ represent the initial free embedding of user $a$ and item $i$, respectively. $\mathcal{N}_r(a)$ is the itemset that user $a$ consumed, $GNN(\cdot)$ denotes a layer-wise graph neural network, and $K$ indicates the number of GCN layers. When the social module is removed, $h_a^K$  simply defaults to $h_a^0$, as the social diffusion layers, which would normally leverage the explicit user-user social graph ($G_s$), are omitted. Therefore, DiffNet (w/o social) relies exclusively on the user-item interaction graph ($G_r$) to generate recommendations.  

From Table~\ref{tab:TRADE-OFF}, we observe that social recommendation models significantly enhance accuracy compared to their non-social counterparts. For instance, SocialMF shows an improvement of approximately 12\% on Yelp and 5\% on Ciao when compared to SocialMF (w/o social). However, in terms of diversity, most social recommendation methods lead to a substantial decrease in system-level diversity. For example, the diversity of DiffNet drops from 40\% to 17\% on Yelp and from 39\% to 28\% on Ciao when compared to its non-social variant. Based on these results, we can conclude that: \textbf{social recommendation methods usually reduce recommendation diversity while improving recommendation accuracy compared to their non-social variants.}

\subsection{Embedding Similarity in Social Recommendation}
\label{sec:lfs}
To further explore the findings presented in Table~\ref{tab:TRADE-OFF} and understand why SocialRS reduces diversity, we analyze the embedding similarities between users and their friends. Following~\cite{sacharidis2020fairness}, we calculate the cosine similarity between the feature vectors of two users, normalizing the values to a range of 0 to 1, where higher values indicate greater similarity.

\begin{figure}[tb]
  \centering
\includegraphics[width=1.0\linewidth]{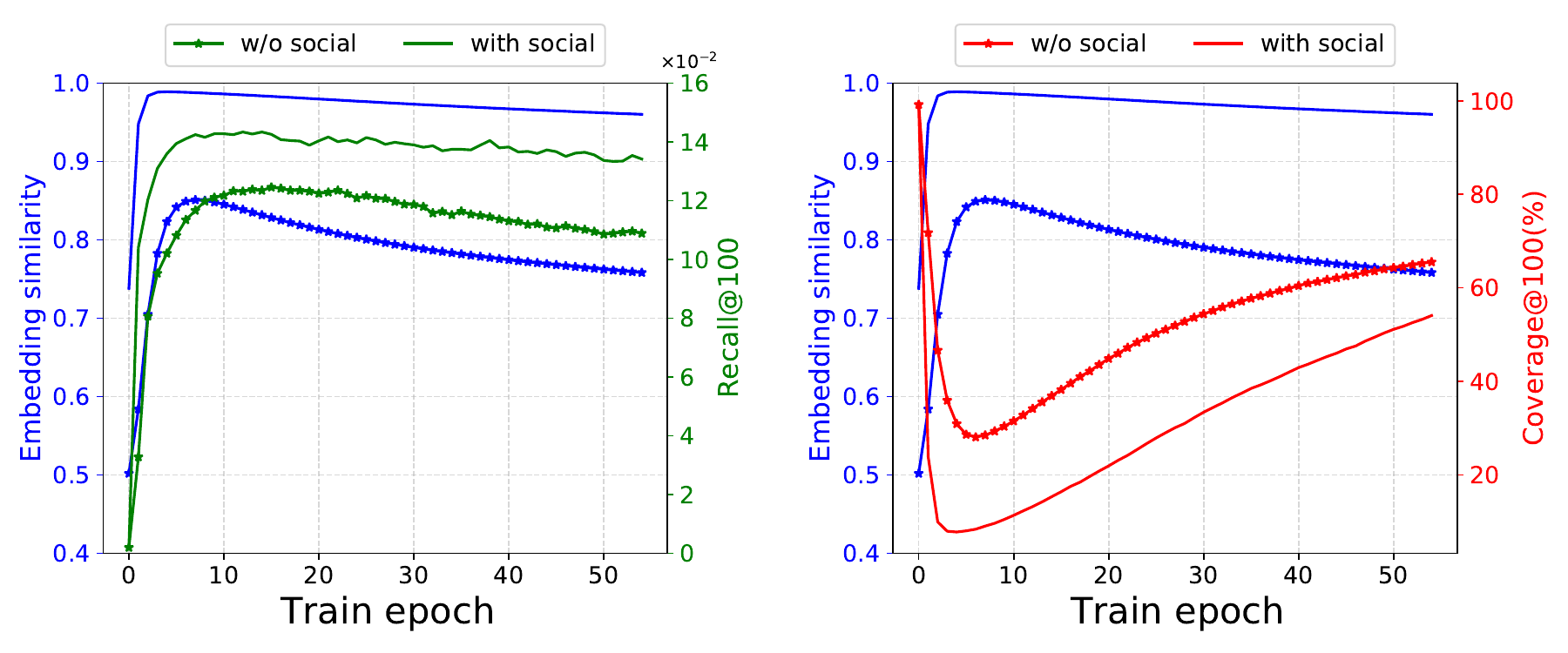}
  \caption{The trends of the user-friend embedding similarity during training.}
  \label{fig:lfs}
\end{figure}

Figure \ref{fig:lfs} shows the changes in embedding similarity, accuracy, and diversity during the training of DiffNet on Yelp dataset. Initially, with random vector initialization, similarity is low, and diversity is high. As training progresses, vector similarities increase, leading to improved accuracy but reduced diversity. In the later stages of training, embedding similarities reach their peak and then begin to decline, while diversity drops to a minimum before gradually increasing again. Meanwhile, accuracy converges toward a stable value. If training continues past convergence, accuracy may decrease as embedding similarities drop and diversity increases. This pattern suggests a negative correlation between embedding similarity and diversity. Additionally, we also observe that social recommendation methods always generate more similar embeddings (higher embedding similarity) between users and their friends compared to their non-social variants.

\section{METHODOLOGY}
\subsection{Overall Framework}
\begin{figure*}[tb]
  \centering  \includegraphics[width=\linewidth]{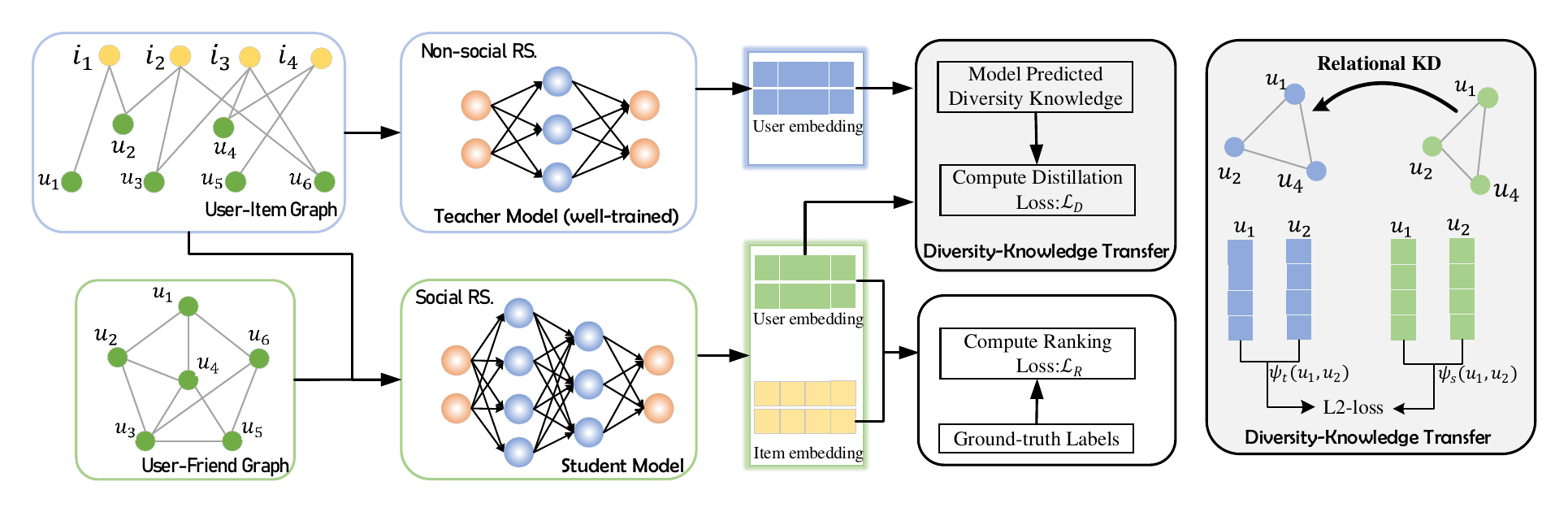}
  \caption{The overall framework of DivSR.}
  \label{fig:MODEL}
\end{figure*}

In this paper, we propose a \textbf{Div}ersified \textbf{S}ocial \textbf{R}ecommendation framework based on knowledge distillation, named \textbf{DivSR}, which achieves a more balanced trade-off between recommendation accuracy and diversity. The overall architecture is shown in Figure~\ref{fig:MODEL}. DivSR is motivated by two key advantages: i) Social RS typically exhibits satisfactory recommendation accuracy, and ii) Non-social RS tends to offer higher recommendation diversity. To leverage these strengths, DivSR introduces a novel approach where a social recommendation model serves as the student model, combining the benefits of both high accuracy and high diversity. The high diversity is derived from a pre-trained teacher model, which is its non-social recommendation counterpart. We design a knowledge transfer module based on relational distillation learning, which facilitates the transfer of structured diversity knowledge from the teacher model to the student model. Specifically, the structured vector similarity between users and their friends is used to capture diversity knowledge, as described in Section~\ref{sec:lfs}, where lower vector similarity corresponds to higher diversity. DivSR optimizes both the recommendation task and the knowledge distillation task simultaneously within a primary and auxiliary learning framework.

\subsection{Teacher Model \& Student Model}
DivSR is a simple and model-agnostic solution that can be easily deployed on existing social recommendation models. Given a social recommendation method, we first pre-train its corresponding non-social variant as the teacher model. using the recommendation loss. Then, we jointly train the social recommendation model as the student model, using both recommendation loss and distillation loss. During the training of the student network, the teacher network has already been fully trained and frozen.

Specifically, we formalize $f_t(\theta_t)$ as a Non-social RS, where $\theta_t$ is the model parameters of the teacher network. Since the teacher model does not account for social relationships, it is provided with a user-item bipartite graph. The teacher model is trained using the Bayesian Personalized Ranking (BPR)~\cite{rendle2012bpr} loss:
\begin{equation}
    \label{eq:T_BPR}
  \mathcal L_R(\theta_t) = \sum_{i \in \mathcal{N}_r(a), j\notin \mathcal{N}_r(a)}-ln\sigma(\tilde{r}_{a,i}(\theta_t)-\tilde{r}_{a,j}(\theta_t))+\lambda {\left \| \theta_t \right \|}^2,
\end{equation}
where $\mathcal{N}_r(a)$ denotes the set of items consumed by user $a$, $\sigma(\cdot)$ is a sigmoid function, and $\lambda$ is a regularization parameter to prevent overfitting. The rating score $\tilde{r}_{a,i}(\theta_t) = (q_i^t)^Tp_a^t$ is determined by the final user embedding $p_a^t$ and item embedding $q_i^t$ produced by the teacher model.

Now, considering the student model $f_s(\theta_s)$, which incorporates both the user-item bipartite graph ($G_r$) and the user-user social graph ($G_s$), we proceed with joint training on both the recommendation task and the knowledge distillation task. The knowledge distillation task will be introduced later and the objective function for the recommendation task is same as Equation~\ref{eq:T_BPR}. The final user/item embeddings generated by the student model for user $a$ and item $i$ are $p_a^s$ and $q_i^s$, respectively, and their dot product forms the rating scoring, $\tilde{r}_{a,i}(\theta_s) = (q_i^s)^Tp_a^s$.

\subsection{Diversity-Knowledge Transfer}
\label{sec:KnowDis}
Knowledge distillation has garnered significant attention for model compression across various domains ~\cite{romero2014fitnets,hinton2015distilling,you2017learning}. It facilitates the transfer of knowledge from a teacher model, which typically has a large capacity, to a student model, thereby preserving comparable performance. Unlike conventional methods that transfer individual outputs from the teacher to the student on a point-wise basis, relational knowledge distillation (RKD)~\cite{park2019relational} is introduced to transfer relational information at the structural level of the outputs.

To enhance recommendation diversity, we leverage the embedding similarity between users and their friends as a diversity indicator. Notably, this similarity operates at a structure-to-structure level, as depicted in Figure~\ref{fig:MODEL}. Consequently, we employ relational knowledge distillation to transfer knowledge from the teacher model to the student model:
\begin{equation}
\label{eq:KD_loss}
  \mathcal L_D = \sum_{(a,b) \in G_s} l_\delta (\psi_T(a, b), \psi_S(a, b)),
\end{equation}
where $l_\delta$ is L2 loss, $\psi(a, b)$ is the angle-wise potentials:
\begin{equation}
\label{eq:phi_loss}
  \psi(a, b) = cos \ \theta = \frac{p_a^Tp_b}{\left\|p_a\right\|\left\|p_b\right\|}.
\end{equation}

The angle-wise distillation loss $\mathcal L_D$ facilitates the transfer of relational information between training example embeddings by penalizing angular discrepancies. Since angles encapsulate higher-order properties compared to distances, they provide a more effective means of transferring relational information, thereby endowing the student model with greater flexibility.

Given the inherent complexity and noise in real-world social relationships, we adopt a pragmatic approach by refraining from computing the similarity between individual user-friend pairs $(a,b)$ in $G_s$. Instead, we compute the average embedding vector of all social neighbors of a user $a$, denoted $p_{af} = mean(p_{\{b\}}), b\in \mathcal{N}_s(a)$, which smooths out feature representations and mitigates noise within the social network.

\subsection{Model Training}
\label{sec:ModelTrain}

The learning process of the student model involves two tasks: the recommendation task and the knowledge distillation task. The overall objective is defined as follows:
\begin{equation}
  \mathcal{L} = \mathcal{L}_R(\theta_s) + \beta \mathcal{L}_D(\theta_s|\theta_t)
  \label{eq:Joint},
\end{equation} 
where $\mathcal{L}_{R}$ represents the recommendation task loss function, $\mathcal{L}_{D}$ denotes the knowledge distillation loss function, and $\beta$ is a hyperparameter that controls the trade-off between the two objectives. A larger value of $\beta$ prioritizes the acquisition of diversity knowledge, while a smaller value emphasizes accuracy. After obtaining the combined representations for all users and items within the student model, we can predict user $a$'s preference for item $i$: $\tilde{r}_{a, i}(\theta_s) = (q_i^s)^Tp_a^s$. 

\begin{table}
\centering
\caption{Statistics of the datasets.}
  \label{tab:DATA}
  \setlength{\tabcolsep}{5pt}
  \resizebox{1.0\linewidth}{!}{
  \begin{tabular}{l|c|c|c|c|c|c}
  \hline
     Dataset &\#Users &\#Items &\#Feedback &\#Relations & Feedback Dens. &Relation Dens.\\
     \hline
      Yelp &17,220 &35,351 &205,529  &143,609 &0.034\% &0.048\%    \\
       Ciao  &6,788  &77,248  &206,143  & 110,383 & 0.039\% &0.239\%\\
      Flickr &8,137   &76,190  &320,775  & 182,078 & 0.050\% &0.275\%  \\
   \hline
    \end{tabular}
}
\end{table}

\section{EXPERIMENTS}

\subsection{Experimental Settings}
\subsubsection{Datasets}
In order to be consistent with previous research ~\cite{wu2019neural,yu2021self}, we conduct experiments on three widely used benchmark datasets. Yelp\footnote{\href{https://www.yelp.com/}{https://www.yelp.com/}} is a popular online location-based social network that allows users to share their experiences. Ciao\footnote{\href{http://www.cse.msu.edu/~tangjili/trust.html}{http://www.cse.msu.edu/~tangjili/trust.html}} is a well-known social networking website where users can rate items, write reviews, and add friends. Flickr\footnote{\href{https://www.flickr.com/}{https://www.flickr.com/}} is an online image-based social sharing platform. The statistics of these datasets are summarized in Table \ref{tab:DATA}. 

\subsubsection{Metrics}
In this work, we adopt four commonly used metrics to measure accuracy and diversity. For accuracy, we utilize Recall@K and NDCG@K~\cite{yu2021self}. Recall computes the fraction of relevant items identified out of all relevant items. NDCG (Normalized Discounted Cumulative Gain) places greater emphasis on higher-ranked resources and incorporates varying relevance levels through different gain values. To measure diversity, we use Coverage@K and Entropy@K, which are frequently applied in diversified recommendation task~\cite{puthiya2016coverage,zheng2021dgcn}. Coverage measures the extent to which a recommendation set covers diverse items from the entire item pool. Entropy assesses the uniformity of item probabilities within the recommendation set. Higher Coverage@K and Entropy@K mean greater diversity. To save space, we only report top-\textit{100} recommendation results, noting that similar conclusions hold for other top-\textit{N} recommendations.
 
\subsubsection{Implementation Details} Our experiments are conducted on NVIDIA V100 GPUs with 32GB memory. For all methods, we refer to the hyperparameter ranges provided in their original papers and perform grid search to identify the optimal set of hyperparameters. We use the Adam optimizer~\cite{kingma2014adam} with a gradient descent-based approach, initializing the learning rate at 0.001. The batch size is set as 2000, and the embedding size is fixed as 64. The $L_2$ regularization parameter $\lambda$ is 0.001. The coefficient $\beta$ for knowledge distillation is searched within the range \{$2.0, 1.0, 0.5,...,1e-4$\}. Each experiment is conducted five times, and the reported results represent the average performance across these runs. Additionally, early stopping is employed to mitigate overfitting.

\begin{table}[tb]
  \caption{Overall performances of five backbone models and DivSR on three datasets. The improvements are calculated between Base and DivSR. All the metrics (except E@100) are percentage numbers with ’\%’ omitted.}
  \label{tab:WHOLE}
  \resizebox{\linewidth}{!}{
  \begin{tabular}{l|l|cc|cc|cc|cc|cc|cc}
  \hline
     &Dataset   &\multicolumn{4}{|c|}{Yelp} &\multicolumn{4}{|c|}{Ciao} &\multicolumn{4}{|c}{Flickr}\\
     \hline
     Backbone& Method &R@100 & N@100 &C@100 & E@100 &R@100 & N@100 &C@100 & E@100  &R@100 & N@100 &C@100 & E@100 \\
     \hline
    \multirow{4}*{TrustMF} & w/o social  &12.513 &3.452 &31.610 &10.792 &12.301 &4.932 &33.478 &10.420 &2.219 &0.797 &40.723 &11.943 \\
    ~ & Base   &13.773 &3.823 &23.276 &10.152 &12.778 &5.081 &21.524 &8.963 &3.242 &1.185 &21.547 &10.230 \\
      ~ & DivSR   &13.721 &3.816 &28.015 &10.329 &12.634 &5.080 &27.245 &9.536 &3.299 &1.186 &25.329 &10.619 \\
      ~ & \textbf{Improve.}   &\textbf{-0.38\%} &\textbf{-0.18\%} &\textbf{20.36\%} &\textbf{1.75\%}  &\textbf{-1.13\%} &\textbf{-0.02\%} &\textbf{26.58\%} &\textbf{6.40\%}  &\textbf{1.76\%} &\textbf{0.08\%} &\textbf{17.55\%} &\textbf{3.80\%} \\
    \hline
    \multirow{4}*{SocialMF} & w/o social      &12.513 &3.452 &31.610 &10.792 &12.301 &4.932 &33.478 &10.420 &2.219 &0.797 &40.723 &11.943 \\
    ~ & Base   &14.077 &3.858 &21.335 &10.391 &12.899 &5.170 &22.215 &9.694 &3.339 &1.211 &27.125 &11.069 \\
      ~ & DivSR  &14.139 &3.897 &35.938 &10.925 &12.942 &5.171 &31.684 &10.201 &3.330 &1.210 &32.424 &11.515 \\
       ~ & \textbf{Improve.}   &\textbf{0.44\%} &\textbf{1.01\%} &\textbf{68.45\%} &\textbf{5.14\%}  &\textbf{0.33\%} &\textbf{0.02\%} &\textbf{42.62\%} &\textbf{5.23\%}  &\textbf{-0.27\%} &\textbf{-0.08\%} &\textbf{19.54\%} &\textbf{4.03\%} \\
        \hline	
    \multirow{4}*{DiffNet} & w/o social      &12.543 &3.395 &39.974 &11.279 &12.519 &4.710 &39.655 &11.385 &2.135 &0.765 &43.654 &12.289 \\
    ~ & Base   &14.136 &3.843 &16.738 &10.225 &12.658 &5.184 &27.560 &10.272 &3.539 &1.263 &32.663 &11.719 \\
      ~ & DivSR   &14.278 &3.868 &26.770 &10.647 &13.133 &5.146 &31.822 &10.784 &3.517 &1.258 &37.160 &11.958 \\
       ~ & \textbf{Improve.}   &\textbf{1.00\%} &\textbf{0.65\%} &\textbf{59.94\%} &\textbf{4.13\%} &\textbf{3.75\%} &\textbf{-0.73\%} &\textbf{15.46\%} &\textbf{4.98\%}  &\textbf{-0.62\%} &\textbf{-0.40\%} &\textbf{13.77\%} &\textbf{2.04\%} \\
    \hline
    \multirow{4}*{DESIGN} & w/o social   &13.553 &3.742 &51.277 &11.898  &14.102 &5.308 &56.267 &12.878  &3.427 &1.213 &32.761 &11.779  \\
    ~ & Base   &14.984 &4.187 &42.228 &11.852  &15.367 &6.001 &24.262 &11.130 &4.127 &1.422 &35.290 &12.828 \\
      ~ & DivSR &15.008 &4.200 &43.371 &11.926 &15.424 &6.007 &26.661 &11.150 &4.172 &1.462 &37.298 &12.979 \\
       ~ & \textbf{Improve.}  &\textbf{0.16\%} &\textbf{0.31\%} &\textbf{2.71\%} &\textbf{0.62\%} &\textbf{0.37\%} &\textbf{0.10\%} &\textbf{9.89\%} &\textbf{0.18\%}  &\textbf{1.09\%} &\textbf{2.81\%} &\textbf{5.69\%} &\textbf{1.18\%} \\
    \hline
    \multirow{4}*{MHCN} & w/o social     &14.518 &4.001 &59.502 &11.813 &15.803 &6.016 &19.746 &10.464 &3.152 &1.174 &22.905 &11.021  \\
    ~ & Base    &15.365 &4.317 &53.273 &11.654 &16.338 &6.504 &22.173 &10.529 &4.574 &1.645 &40.205 &12.173 \\
      ~ & DivSR   &15.323 &4.306 &55.903 &11.674 &16.336 &6.441 &25.572 &10.536 &4.579 &1.633 &43.757 &12.268 \\
       ~ & \textbf{Improve.}  &\textbf{-0.27\%} &\textbf{-0.25\%} &\textbf{4.94\%} &\textbf{0.17\%}  &\textbf{-0.01\%} &\textbf{-0.97\%} &\textbf{15.33\%} &\textbf{0.07\%}  &\textbf{0.11\%} &\textbf{-0.73\%} &\textbf{8.83\%} &\textbf{0.78\%} \\
\hline
\end{tabular}
}
\end{table}

\subsection{Main Results with Various Backbone Models}

\subsubsection{Backbones.} Since DivSR is model-agnostic, we evaluate its performance with several representative social recommender systems. TrustMF~\cite{yang2016social} employs matrix factorization (MF) to embed users into low-dimensional spaces. SocialMF~\cite{jamali2010matrix} is a regularization-based social recommendation model that constrains users' latent vectors to be close to those of their social neighbors. DiffNet~\cite{wu2019neural} utilizes graph convolutional networks to capture dynamic social diffusion in social graphs. MHCN~\cite{yu2021self} enhances social recommendation through self-supervised learning (SSL) on motif-induced hypergraphs. DESIGN~\cite{tao2022revisiting} performs statistical data analyses to gain deeper insights into the theory of social influence.

\subsubsection{Results.} We train both the base social models and their DivSR counterparts on three datasets. The overall recommendation results are presented in Table~\ref{tab:WHOLE}.

From the results, we observe that social recommendation methods usually reduce recommendation diversity while improving recommendation accuracy compared to their non-social counterparts (except DESIGN and MHCN on the Flickr dataset). For DESIGN and MHCN on Flickr, we guess that the effect of social connections on user preferences is less pronounced. These models leverage social relationships as supplementary information for self-supervised learning, rather than directly modeling user preferences based on social influence. Regarding accuracy, GNNs-based methods consistently outperform MF-based methods (\textit{i.e.} DiffNet \textit{vs.} TrustMF), which can be attributed to the superior capability of GNNs in modeling graph data. SSL-enhanced methods prove to be more effective than methods without SSL (\textit{i.e.} MHCN \textit{vs.} DiffNet), highlighting the effectiveness of self-supervised learning in enhancing performance. In terms of diversity, GNN-based methods tend to generate more diverse recommendations compared to MF-based methods (\textit{i.e.} DESIGN \textit{vs.} SocialMF), likely due to their ability to incorporate a broader range of neighbors for users/items. 

DivSR promotes diversity without significantly sacrificing accuracy and, in some cases, even improves it compared to these social recommendation backbones. For example, DivSR boosts diversity from 23\% to 33\% on average across the three datasets towards SocialMF, with accuracy fluctuating by less than 1\%. When compared to GCN-based backbones, DivSR achieves a more notable improvement in diversity over MF-based backbones. Furthermore, DivSR typically demonstrates a higher coverage while maintaining a comparable recall score, indicating a better balance between accuracy and diversity. It's noteworthy that DivSR may occasionally show slightly lower diversity than Social RS (w/o social), but the accuracy of the latter is significantly lower. These findings underscore the effectiveness of the knowledge distillation module in enhancing recommendation diversity while maintaining high accuracy.

\begin{figure}[tb]
  \centering
  \includegraphics[width=0.75\linewidth]{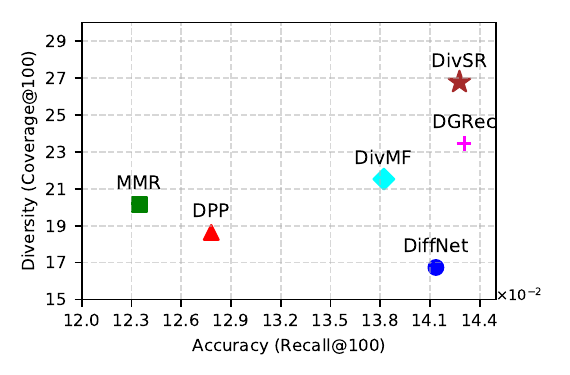}
  \caption{Accuracy-Diversity trade-off comparison on Yelp dataset.}
  \label{fig:baseline}
\end{figure}

\subsection{Comparison with Diversified Models}
\subsubsection{Baselines.} We conduct experiments to compare DivSR with several diversified methods. MMR~\cite{carbonell1998use} leverages marginal relevance to balance relevance and diversity in recommendation lists. DPP~\cite{chen2018fast} is a probabilistic model that is widely adopted for generating diversified recommendations. DivMF~\cite{kim2023diversely} regularizes the score matrix of an MF model to maximize the coverage of top-\textit{k} recommendation lists. DGRec~\cite{yang2023dgrec} designs a submodular function to select a diversified subset of neighbors, thereby enhancing diversity.

\subsubsection{Results.} As shown in Figure~\ref{fig:baseline}, two re-ranking methods, MMR and DPP, prioritize diversity by generating more varied recommendations. However, this comes at the cost of a significant drop in recommendation accuracy, highlighting their inability to effectively balance the accuracy-diversity trade-off. In contrast, DivSR positions itself in the upper-right quadrant, demonstrating its capability to achieve the optimal trade-off between accuracy and diversity. When compared to DGRec, DivSR offers a notable increase in diversity while incurring only a minor sacrifice in accuracy.

\subsection{Teacher Model Choice Analysis}
\label{sec:KNOWDISSANALYSIS}
In this section, we explore two distinct approaches for providing the supervised signal to train the student model. \textbf{Uns.:} It removes the teacher model and directly minimizes the user-friend similarity in the student model, $\mathcal L_D = \sum_{(a,b) \in G_s} {\left\|\psi_S(a, b)\right\|}_2$, akin to an unsupervised training approach. \textbf{DivSR(M):} It selects the model with the highest diversity from five non-social models as the teacher model. Specifically, we use MHCN (w/o social) as the sole teacher model for Yelp, DESIGN (w/o social) for Ciao, and DiffNet (w/o social) for Flickr. 

\begin{table}[tb]
\centering
  \caption{Performance comparison between different teacher strategies.}
  \setlength{\tabcolsep}{10pt}
  \label{tab:VARIANT}
  \resizebox{1.0\linewidth}{!}{
  \begin{tabular}{l|cc|cc|cc}
  \hline
     Dataset   &\multicolumn{2}{|c|}{Yelp} &\multicolumn{2}{|c|}{Ciao} &\multicolumn{2}{|c}{Flickr}\\
     \hline
     Method &R@100  &C@100  &R@100  &C@100  &R@100  &C@100\\
   \hline
    SocialMF  &14.077 &21.335 &12.899 &22.215 &3.339 &27.125 \\
    \ \ Uns.  &13.802 &24.469 &12.824 &29.078 &\underline{3.299} &\underline{32.458} \\
    \ \  DivSR(M)  &\underline{14.059} &\underline{34.698} &\textbf{13.151} &\textbf{31.975} &3.307 &31.890 \\
    \ \  DivSR   &\textbf{14.139} &\textbf{35.938} &\underline{12.942} &\underline{31.684} &\textbf{3.330} &\textbf{32.424}  \\
      \hline
     DiffNet  &14.136 &16.738 &12.658 &27.560 &3.539 &32.663 \\
    \ \ Uns.   &13.685 &22.936 &12.828 &31.254 &2.298 &33.053 \\
    \ \ DivSR(M)  &\underline{13.829} &\underline{25.270} &\textbf{13.245} &\textbf{32.013} &\textbf{3.517} &\textbf{37.160} \\
     \ \ DivSR  &\textbf{14.278} &\textbf{26.770} &\underline{13.133} &\underline{31.822} &\textbf{3.517} &\textbf{37.160}  \\
\hline
\end{tabular}
}
\end{table}

The results presented in Table~\ref{tab:VARIANT} compare the performance of different teacher model choices. Firstly, the unsupervised method demonstrates that optimizing user features for diversity by directly minimizing user-friend similarity proves effective. Although this may lead to a decrease in accuracy, it frequently results in improved diversity. Secondly, utilizing the supervised signal from the teacher model to guide the transfer of diversity knowledge yields notable benefits for overall performance. Both DivSR and DivSR (M) consistently achieve superior trade-off results compared to the unsupervised method in most cases. Finally, DivSR generally outperforms DivSR (M), consistently ranking among the top-\textit{2} in comparisons. These findings suggest that carefully optimizing user-friend similarity can significantly enhance the accuracy-diversity balance, and that selecting an appropriate teacher model can further enhance performance.

\begin{figure}[tb]
    \centering
    \begin{minipage}{1.0\textwidth}
\includegraphics[width=\textwidth]{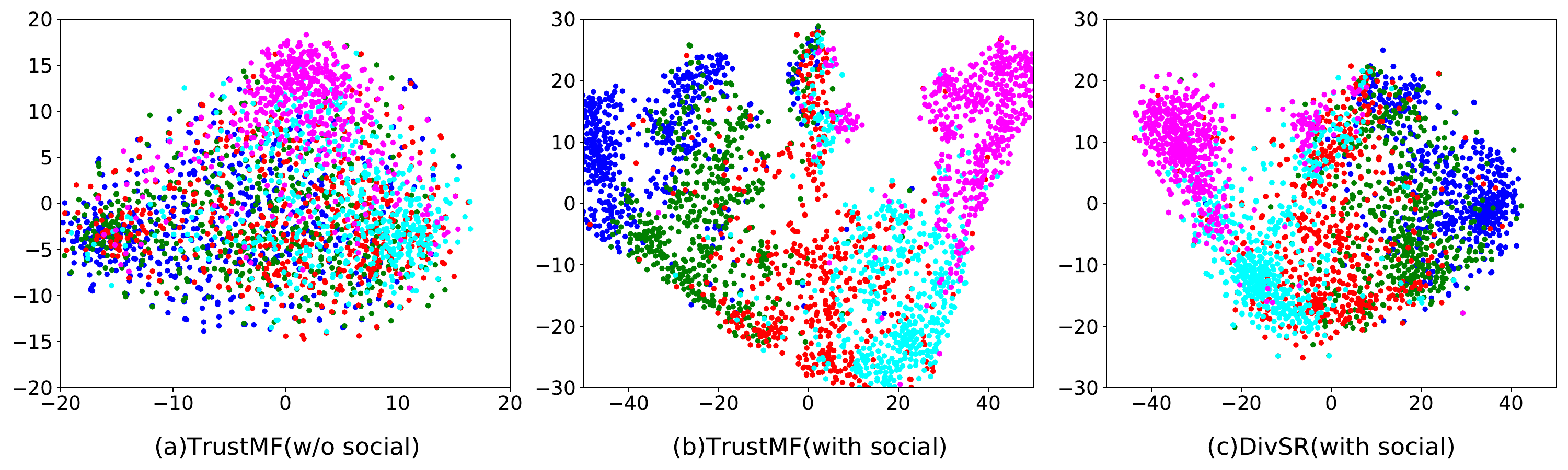} 
        \caption{User embeddings visualization.}
        \label{fig:tsne}
    \end{minipage}
    \begin{minipage}{0.8\textwidth}
\includegraphics[width=\textwidth]{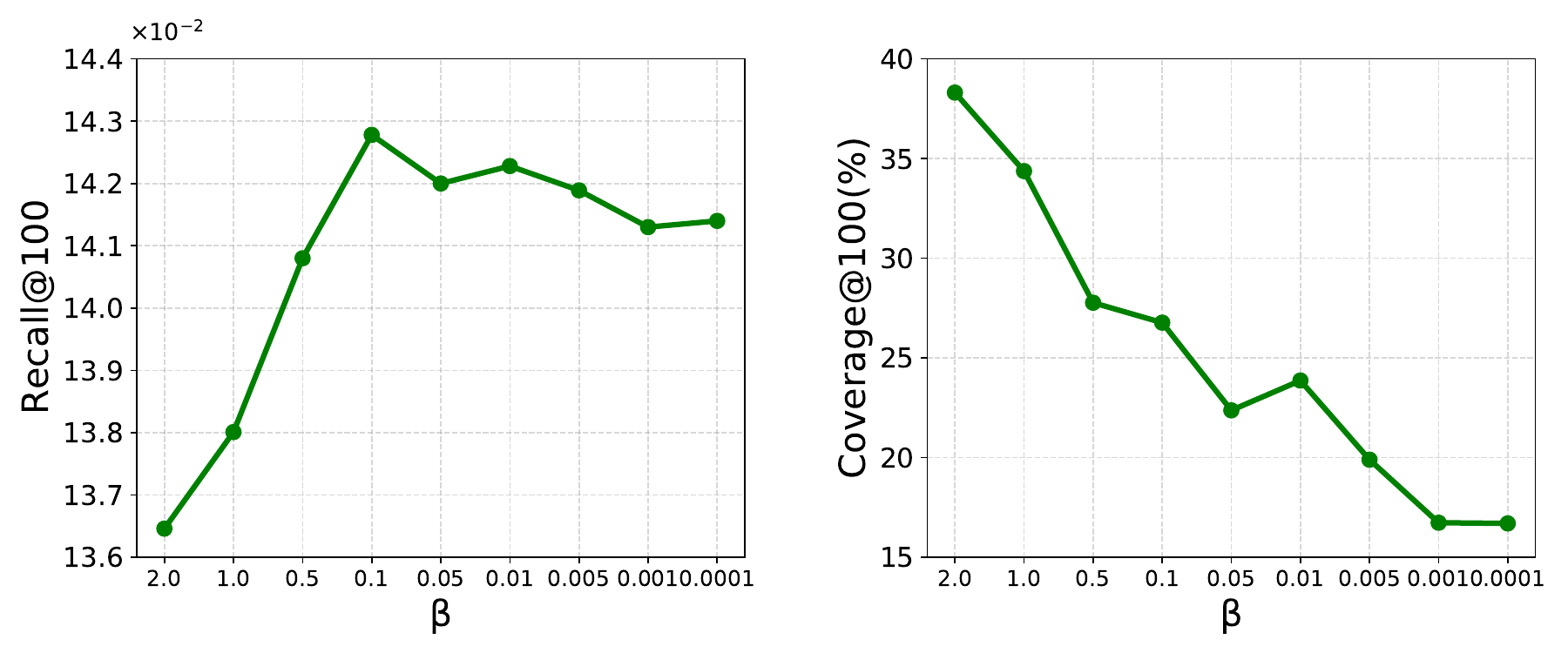} 
        \caption{Parameter sensitivity.}
        \label{fig:belta}
    \end{minipage}
\end{figure}

\subsection{Effect of Knowledge Distillation}
To gain a deeper understanding of the impact of the knowledge distillation mechanism in DivSR, we present a visualization of user embeddings in Figure~\ref{fig:tsne}. Using the Louvain algorithm~\cite{blondel2008fast}, we detect communities within the Flickr dataset. We then randomly sample 2,000 users from the top five communities and apply t-SNE~\cite{van2008visualizing} to visualize their embeddings.

From the visualization, we observe that TrustMF (w/o social) produces relatively disorganized user representations, with no clear community aggregation. In contrast, TrustMF (with social) displays distinct community segregation, where users within the same community are closely grouped, while those in different communities remain well-separated. This pronounced segregation can limit user engagement and hinder content diversity. However, DivSR strikes a balance by exhibiting some degree of community aggregation while reducing excessive community segregation, thus fostering greater inter-community interaction. These results demonstrate that the distillation mechanism effectively encourages the student model to learn low similarity between users, as seen in non-social recommendation systems, underscoring the efficacy of DivSR.

\subsection{Parameter Sensitivity}
In this section, we investigate the impact of varying values of $\beta$ on the trade-off between accuracy and diversity. As shown in Figure~\ref{fig:belta}, as $\beta$ decreases, recall gradually increases, reaching its peak at a value of 0.1, before declining toward convergence. In contrast, diversity steadily decreases until it stabilizes. As $\beta$ decreases, the model assigns less weight to diversity learning, leading to a reduction in diversity and an increase in accuracy. When $\beta$ is within an optimal range (\textit{i.e.} 0.05-0.1), the model achieves a favorable balance, simultaneously maintaining high accuracy and high diversity. However, as $\beta$ approaches very small values, the constraint on diversity learning diminishes significantly, causing the model to degenerate into a purely social recommendation model, with both accuracy and diversity converging to the performance of the baseline model. The coefficient $\beta$ thus plays a critical role in balancing the primary recommendation task and the auxiliary knowledge distillation task.

\section{CONCLUSION}
In this work, we examine the trade-off between recommendation accuracy and diversity in social recommendation models. Through empirical analysis, we observe that many existing social recommendation methods tend to reduce diversity while improving accuracy compared to their non-social variants. To address this challenge, we propose DivSR, a simple yet model-agnostic framework that can be seamlessly integrated into existing social recommendation systems. DivSR leverages relational knowledge distillation techniques to transfer high-diversity structured knowledge from non-social models to social recommendation models. Our experiments demonstrate that DivSR significantly enhances diversity without substantially compromising accuracy. Moreover, DivSR achieves a superior accuracy-diversity trade-off compared to several diversified models, effectively mitigating the inherent tension between accuracy and diversity.

%
%
%
\bibliographystyle{main}
\bibliography{main}

\end{document}